\documentclass{llncs}

\usepackage{mathtools}
\usepackage{todonotes}
\usepackage{lastpage}
\usepackage{eurosym}

\newcommand{\kreis}[1]{\unitlength1ex\begin{picture}(2.5,2.5)%
\put(0.75,0.75){\circle{2.5}}\put(0.75,0.75){\makebox(0,0){#1}}\end{picture}}

\usepackage{tikz}

\begin{document}

\mainmatter 
\title{Visualizing Multiple Process Attributes in one 3D Process Representation}

\author{Manuel Gall, Stefanie Rinderle-Ma}
\institute{University of Vienna, Faculty of Computer Science, Vienna, Austria\\
\email{manuel.gall@univie.ac.at, stefanie.rinderle-ma@univie.ac.at}}
\maketitle

\begin{abstract}
Business process models are usually visualized using 2D representations. However, multiple attributes contained in the models such as time, data, and resources can quickly lead to cluttered and complex representations. To address these challenges, this paper proposes techniques utilizing the 3D space (e.g., visualizing swim lanes as third dimension). All techniques are implemented in a 3D process viewer. On top of showing the feasibility of the proposed techniques, the 3D process viewer served as live demonstration after which 42 participants completed a survey. The survey results support that 3D representations are well-suited to convey information on multiple attributes in business process models.

\keywords{Process Model, Process Attributes, 3D Representation, Virtual Reality}
\end{abstract}

\section{Introduction}

%General
Companies are sometimes challenged by understanding their own processes as they comprise data from various sources such as databases, services, and real-time information. For their core processes companies use business process models in notations such as BPMN \cite{Recker2010}. These models might contain a huge amount of data and can be very complex in their representation. In order to deal with such complex representations, views have proven to be helpful for reducing the complexity by restricting the provided information \cite{smirnov2012business}. However, by using views relations between attributes cannot be identified.
%Why 3D
To overcome the gap of knowledge representation this paper introduces a 3D representation (3DViz). 3D representations are better suited for presenting more information compared to 2D representations \cite{Schoenhage2000}. Some fundamental approaches for 3D processes representations have been proposed \cite{Betz3DRepOfBusModel,BrownConceptModelingVirtuWorld,Schoenhage2000}. However, representing 3D processes has not moved as far as other disciplines where 3D representations are common nowadays such as bioinformatics, (3D protein modeling) and mathematics (complex simulations). There are various scenarios where 3D processes representations could be of use like, business process analysis, social modeling, and runtime representation.

%Compared to other domains 3D models have not moved as far in the representation of processes a

%Until now using 3D models in the domain of business processes has not moved as far as it did in other domains i.e. Bioinformatics (3D protein modeling) and Mathematics (complex simulations). There are various scenarios where 3D processes representations could be of use like, business process analysis, social modeling and runtime representation.

%Shortcommings
In face of those broad applications one might ask why 3D representations are not yet established for business processes? When first works for 3D process representations were released the computers were not able to render huge models \cite{Schoenhage2000}. Furthermore there is a representation gap when visualizing a 3D model on a 2D monitor \cite{Schoenhage2000}.

\begin{figure}[tbh!]
	\centering
	\includegraphics[width=1\linewidth]{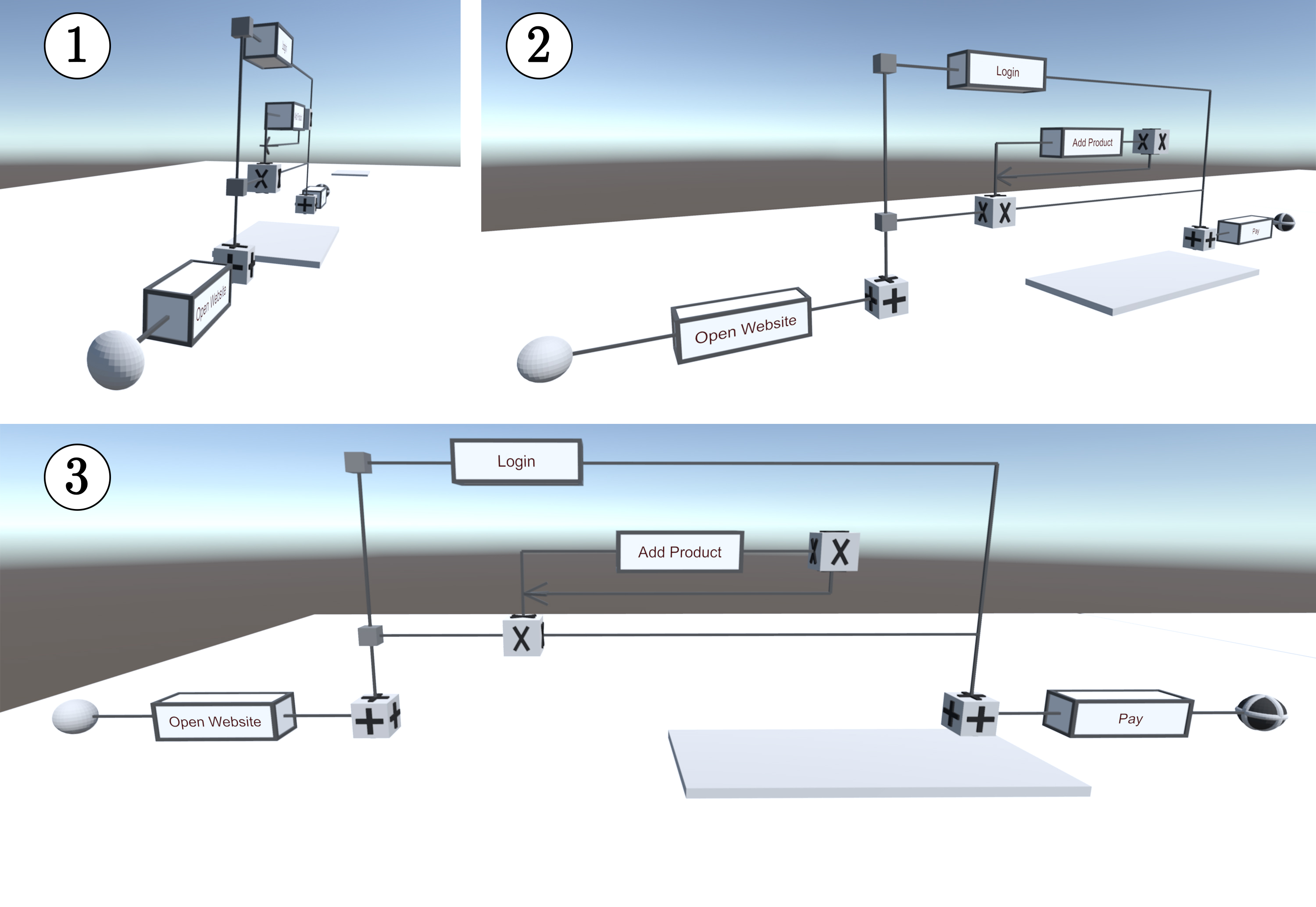}
	\caption{An order process viewed through a VR device from various angles.}
	\label{fig:3DWalkAround}
	\vspace{-0.2cm}
\end{figure}

However, in the last decade the computing power has increased and therefore rendering huge models is not a challenge any more. In recent literature \cite{butscher2018clusters} the second shortcoming (representation gap) is solved by using new devices called head-mounted displays (HMDs) such as virtual reality (VR) and augmented reality (AR) to explore and interact with 3D representations.
%When exploring processes with such devices one can view a process from various angles by moving the head and walking around a process.
Figure \ref{fig:3DWalkAround} shows a simple ordering process visualized on a desk containing four \emph{nodes}, a \emph{parallel} and a \emph{loop}. A person wearing a VR device can literally walk around the process. The perspective on the process is changed by an angle of approximately 30 degree per image from \:\kreis{1} to \:\kreis{3}.

Based on the technological progress (computing power and HMDs) we can argue that the foundation for 3D process representation is available. Within this paper we want to focus on how 3D representations can support the analysis of processes models. While 2D approaches use different views for attribute visualization on a process a 3D representation can contain multiple attributes within one representation.

Specifically we want to look into the following research questions:\\
RQ1: How can multiple process attributes be integrated in one representation?\\
RQ2: How can various data types and resources be integrated in a multi-perspective representation?\\
RQ3: How to go from a 2D towards a 3D representation? \\
RQ4: How to improve navigation and orientation within a virtual setting? \\
%RQ5: Compared to 2D can we use 3D to create models with less overlapping edges? \\

These questions impose various challenges, i.e., avoiding cluttered representations, dealing with different data types, and conflicting visual representation. 
% how to avoid cluttered representations
% -> by reducing edges (direct mapping on each node)

% dealing with different data formats
% -> 

We use the design science research methodology \cite{Wier2015} to answer these questions and tackle the challenges. By using distinct visual styles as well as applying mapping techniques on data types we allow for integrating multiple perspectives into one representation. Section \ref{sec:RelWork} discusses related work for attribute representation and 3D representations. In Section, \ref{sec:Fundamentals} a fundamental 3D process representation framework (3DViz) is introduced. 3DViz is refined in Section \ref{sec:Contribution} to allow for visualizing multiple attributes within one representation. The created framework is evaluated in Section \ref{sec:Evaluation} with a survey and an applicability analysis.

%In accordance to the methodology we designed an artifact for 3D model representation and evaluated this artifact with a user study. 

%What we do
%This paper introduces a fundamental 3D representation of a process model and further shows how multiple views on a process can be 

%fist use case on how such a 3D model can be used.

\section{Related Work}
\label{sec:RelWork}

%2D Ansätze zur prozesssichtvereinigung einbringen und ihre negativen sachen (zu unübersichtlich) bei vielen elementen und oder beim umschalten zwischen ansichten gehen die zusammenhänge verloren.

%Work that proposes a different method to solve the same problem.
%Work that uses the same proposed method to solve a different problem.
%A method that is similar to your method that solves a relatively similar problem.
%A discussion of a set of related problems that covers your problem domain.

Within business processes various attributes such as resource, data and time have been identified \cite{vanDerAalst2011process}. Business process modeling languages have been extended to support the creation of individual process attribute visualizations for various stakeholders \cite{LAROSA2011313}. Such visualizations are usually limited to represent one process attribute \cite{finkelstein1992viewpoints,BeckerGuid}. As our contribution is to highlight relations between attributes, approaches that utilize abstraction techniques \cite{smirnov2012business} such as merging activities are not within the scope of the paper.

Besides 2D representations, 3D representations have been used for representing such single attribute process models \cite{Betz3DRepOfBusModel}. Other 3D representation approaches within the business process domain allow for modeling of 3D process \cite{Brown_CollaborativeModeling_2011,BrownConceptModelingVirtuWorld}. When compared to 2D representations such 3D representations allow for incorporating more information into one representation \cite{Schoenhage2000}. However, these approaches do not consider representing multiple attributes within one representation.

Additionally these approaches use monitors to visualize 3D representations which is not suitable \cite{Schoenhage2000}. Other domains use HMDs to visualize 3D representations \cite{butscher2018clusters}. With our research we want to visualize multiple attributes within one 3D representation.

\section{Fundamentals of the 3D Approach}
\label{sec:Fundamentals}
Before diving into the contribution we lay out the fundamentals for a 3D process representation framework (3DViz) based on existing work. In this section we discuss the basic representation as well as interaction possibilities with the model.

\subsection{Engine}
Previous work uses Java3D \cite{Schoenhage2000} and Second Life \cite{Brown_CollaborativeModeling_2011} for 3D process representation. However, when compared to how 3D Engines evolved these approaches seem to be visually outdated. In recent years engines for creating massive 3D worlds went from closed systems accessible for a hand full of developers to broad open communities \cite{SMITH2009559}. There are more then 600 engines currently available \cite{GameEngines}. We need an engine that satisfies the following criteria. In order to visualize many instances or big processes the engine has to be able to draw many objects. Furthermore the engine shall allow communication to other services e.g., a process engine. As the proposed framework should not only visualize elements on screen but on other devices as well e.g., on augmented reality glasses and virtual reality glasses, an engine that supports such devices will be selected. Based on these criteria one can see that there have been recent developments in the area of 3D representations e.g., rendering of large amounts of objects, AR, and VR. Therefore we are looking for an engine that constantly aligns to the use of future technology.

Big engines e.g., Amazon Lumberyard, Cry Engine, Godot Engine, Unity and Unreal Engine support most of the criteria. As the authors have previous experience with Unity, Unity is utilized for creating the 3D process representation framework 3DViz.

\subsection{Creating a 3D Process Model}
\label{sec:3DProcessModel}
%Intro
In this section every aspect ranging from reading process description files towards visualizing a 3D process model is covered.

%Reading a file and creating a model
\subsubsection{File Handling} We use XML files from the CPEE \cite{CPEE} process modeling/execution engine as input. These files are not tailored towards a 3D representation as most of the formats out there and this is the challenge we want to tackle. We want to use input files that are not tailored towards 3D to show that those can be used in a 3D representation ($\mapsto$ RQ 3). %and because there is a wide repository of files used for 2D representation.
While reading a file each element e.g., task, loop and parallel is internally stored as an object. In the following we call these objects \emph{Nodes} and store them within a list. A \emph{Node} consists of the following attributes:

\begin{itemize}
\item \emph{Types} express the elements control flow behavoiur e.g., loop and parallel.
\item \emph{Labels} are used for tasks as well as conditions e.g., XOR conditions. 
\item \emph{Id's} are unique identifier that allow for \emph{Node} identification. 
\item \emph{Nodes Before} is a set of \emph{Nodes} that occur immediately before the current \emph{Node}.
\item \emph{Nodes After} is a set of \emph{Nodes} that occur immediately after the current \emph{Node}.
\item \emph{Arguments} comprise all the additional values stored within the XML file and is designed to be a flexible structure as not all files contain all kinds of arguments. Examples for arguments are the time needed to execute a \emph{Node}, role executing a \emph{Node}, costs when executing a \emph{Node}, services used when the \emph{Node} is executed and many more.
\end{itemize}

%elements
\subsubsection{Visual Elements}
\begin{figure}[tbh!]
	\centering
	\includegraphics[width=0.8\linewidth]{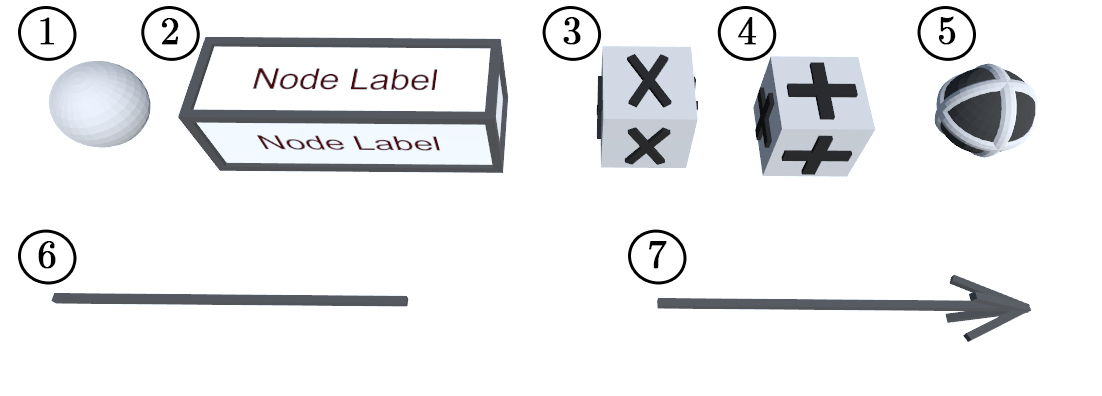}
	\caption{3D elements available within the 3DViz framework.}
	\label{fig:3DElements}
\end{figure}

Every type stored within the \emph{Nodes} is represented by one specific 3D visual element. These 3D elements are influenced by BPMN as it is the defacto standard for modeling business processes \cite{Recker2010}. Fig. \ref{fig:3DElements} shows all the representation elements in the 3DViz framework containing  \:\kreis{1} Start Event, \:\kreis{2} Task, \:\kreis{3} Parallel Split/Join, \:\kreis{4} Xor Split/Join and Loop, \:\kreis{5} End Event, \:\kreis{6} Edge and \:\kreis{7} Edge with arrow. These visual elements were influenced by BPMN and refined by design principles \cite{Moody}.

%Drawing the control Flow
\subsubsection{Drawing the Control Flow}
%Based on the internal structure the control flow perspective is visualized.
Creating the control flow is achieved by iterating through a list of \emph{Nodes} and drawing the appropriate visual element determined by the \emph{Type} attribute. There are \emph{Types} leading to a more complex behaviour when drawing the process model e.g., loops, parallel and XOR. When handling these special \emph{Types} we have to be aware that they can influence the positioning of other \emph{Types}. For example, a parallel branch could consist of an XOR with multiple branches therefore the height positioning of the next parallel branch has to be calculated based on the maximum height of the last branch. 

Figure \ref{fig:3DWalkAround} depicts an example were the height positioning of the second parallel branch is calculated based on the height of the branch below. This is achieved by drawing the first branch. Then the maximum height of the branch is calculated. Based on the maximum height the second branch is drawn.

%\begin{figure}[tbh!]
%	\centering
%	\includegraphics[width=1\linewidth]{images/Perspectives.jpg}
%	\caption{Perspectives on a process.}
%	\label{fig:3DPerspective}
%\end{figure}

%Drawbacks

%Our Framework currently only supports the visualization of a process model. Modeling of a process is currently not in scope.

\subsection{Virtual and Augmented Reality}
As literature suggests viewing a 3D model on a monitor might not be the appropriate way to transport the visual fidelity introduced with a 3D model \cite{Schoenhage2000}. Therefore we opted for extending our representation beyond desktop monitors. We allow to view a process model with various devices. So far we have implemented the viewing on a monitor and an HTC Vive VR device. Viewing the model in AR is implemented but not tested due to the lack of a device.

\subsection{Interaction with the Model}
When interacting with a model there are common actions every representation shall support. These are \emph{zooming}, \emph{panning}, \emph{rotating} and \emph{click} actions \cite{Schoenhage2000}. Based on the device used for viewing a process model we implemented different interaction possibilities.

\subsubsection{Monitor}
When a monitor is used a mouse interacts with the process model. We allow for \emph{zooming} by using the scroll wheel, \emph{panning} by holding the left mouse button and moving the mouse, \emph{rotating} by holding the right mouse button and moving the mouse and \emph{click} actions by left clicking on a \emph{Node}. Click actions are used to get more details on a specific \emph{Node} e.g., visualizing id, responsible roles and other arguments.

\subsubsection{VR Device}
A VR device allows for the same interactions as the monitor/mouse combination. However, these interactions are implemented in a different way. \emph{Zooming}, \emph{panning} and \emph{rotating} are direct results of the head movement observed by the VR device. For example, when a certain \emph{Node} is interesting leaning towards the \emph{Node} will \emph{zoom} in. The \emph{click} action is realized by using VR controllers. When a controller is near a \emph{node} a \emph{click} selects the node and shows the same information as a mouse click will show.

\subsection{Using a 3D Scene}
All the figures so far have something in common, they highlight the process by only showing process relevant information. However, within the framework the representation is different. 

\begin{figure}[tbh!]
	\centering
	\includegraphics[width=0.8\linewidth]{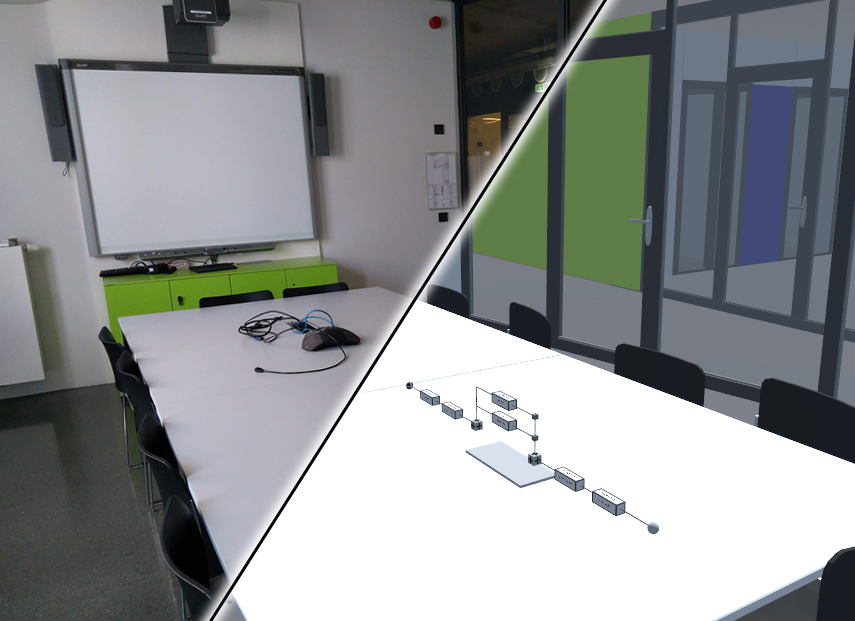}
	\caption{The left side depicts a picture of a real world room, the right side a virtual conference room.}
	\label{fig:RealVirtualWorld}
\end{figure}

We digitalized one of our conference rooms. To be as accurate as possible, every objects real world size was measured. Afterwards all the objects were created with a 3D modelling tool. This approach allowed to have a 1:1 scale between real and virtual world. In fact if the real and virtual world objects are aligned the same and an object is touched in the virtual world one would feel the real world object. From a pure visual point of view the created scene can be seen as a digital twin of our conference room. The focus of the scene is on top of two tables in the middle of the conference room. On top of these tables the 3D process models are visualized. Fig. \ref{fig:RealVirtualWorld} visualizes the real world conference room on the left and the virtual conference room with a process model on the right. 

Using a 3D scene aids the navigation and orientation in a virtual environment ($\mapsto$ RQ 4). For example, when loosing focus of the model by moving the head or rotating with the mouse one still knows where to find the model as there are points for orientation within the scene. Such a scene can be slightly compared to a grid or ledger lines within a 2D representation. %Fig. XY from the evaluation section depicts some models with a surrounding scene. 

We suggest to use scenes for all 3D representations. Scenes can be used in single user settings such as business process analysis and runtime evaluation. In a multi user setting like collaborative modeling scenes can help to improve communication. We want to point out that a scene around the process only makes sense when wearing a VR device or using a monitor.

\iffalse
3) Using a 3D scene around the processes helps interaction/navigation.
Former approaches just show a model, we assumed that a "natural" setting around the model will ease the navigation and help interaction with the model. Besides helping the navigation for experienced users a scene can help reducing the time needed to learn how to navigate within the model.

4) Less overlapping edges
\fi

%However we added an other interaction used for quickly moving inside a room called \emph{teleport}. 

%3) Interaction

%Introduce/Explain Requirements by \cite{Schoenhage2000}, As of interacting with our 3D environment we opted for things used in a 2D environment but can be used in a 3D environemnt i.e., zooming, panning and clicking

% Explain that interactions are currently for two input devices, VR Controller and Mouse

%outro 
In conclusion these fundamentals allow us to draw a 3D process model read from a file. This model can be explored with various devices and allows some interactions.

\section{Attribute Visualization}
\label{sec:Contribution}
So far 3D representations have been mainly concerned with control flow visualization. The 3DViz representation presented in this paper extends the control flow view (basic 3DViz control flow elements are presented in Section \ref{sec:3DProcessModel}) by visualizing multiple attributes within one representation. This particularly enables the detection of relationships between attributes.

\subsection{Example Process}
\begin{figure}[htb!]
	\centering
	\includegraphics[width=1\linewidth]{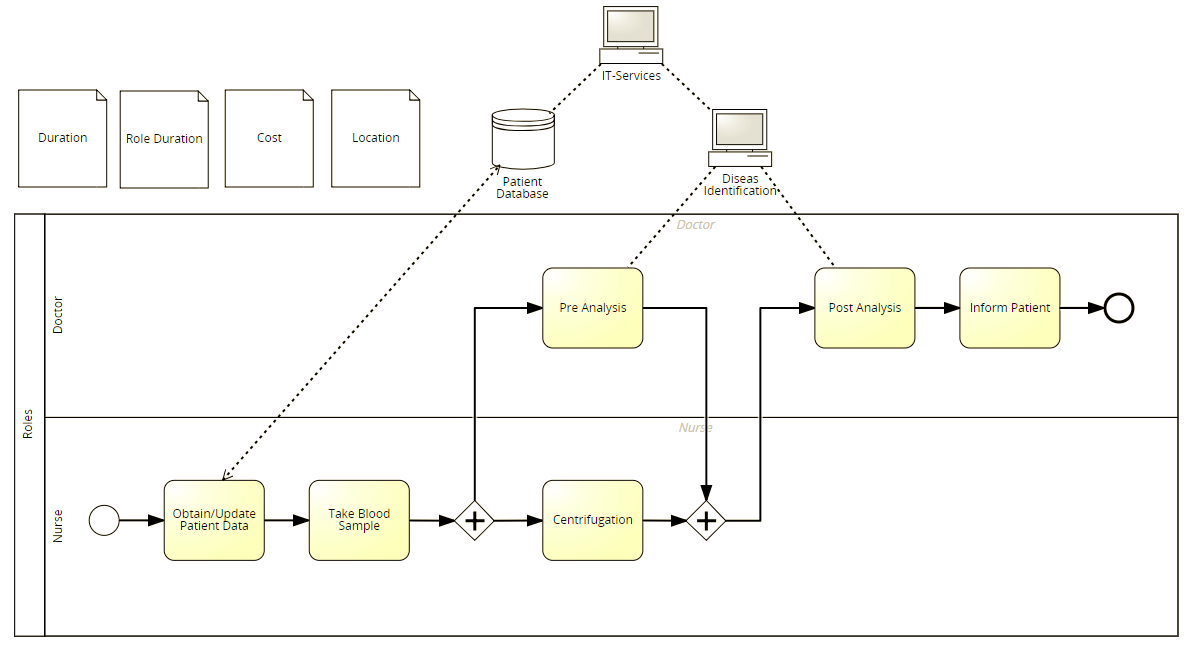}
	\caption{An example process created with \emph{Signavio} represented as BPMN model and containing different attributes.}
	\label{fig:ExampleProcess}
\end{figure}

For the remainder of the paper we will use the process depicted in Fig. \ref{fig:ExampleProcess}. The process describes a common scenario from the care domain for non stationary patients. A patients blood is analyzed to detect a possible disease the patient is suffering from. Fig. \ref{fig:ExampleProcess} shows the process containing these elements: two roles \emph{Nurse} and \emph{Doctor}, six \emph{Nodes}, two \emph{IT-Services} and four data elements \emph{Duration}, \emph{Role Duration}, \emph{Cost} and \emph{Location}. Every \emph{Nodes} is connected to each of the 4 data elements. However in order to keep a readable figure those \emph{Nodes} and data elements are not connected in Fig. \ref{fig:ExampleProcess} as the edges would span across other \emph{Nodes}. The data elements possess different units and data types \emph{Duration} and \emph{Role Duration} in minutes and \emph{Cost} in \euro \: are numerical data types while \emph{Location} is a string. Some tasks are linked to \emph{IT-Services}. These services are identified by their URL and represented as string. A \emph{Node} must not contain all the available arguments. For example, the \emph{Node} \emph{Centrifugation} includes the arguments for all four data elements, the role \emph{Nurse} and no \emph{IT-Service}. Table \ref{tab:exampleValues} comprises the provided values for the data elements. These are estimates and shall support the understanding of the visuals shown in the next sections.

%More detailed describtion:
% The patient comes in and is given a digital form to fill out. Afterwards a blood sample is taken by a nurse. A portion of the blood sample is put into the centrifuge by the nurse and an other portion is analyzed by the doctor. After both are finished the doctor analysis the centrifuged blood sample and informs the patient.

%Role duration = how long the role is involved in this task, this can be different from the tasks duration.

\begin{table}[tbh!]
\begin{tabular}{|l|l|l|l|l|l|l|}
\hline
    & \begin{tabular}[c]{@{}l@{}}Obtain/Update\\ Patient Data\end{tabular} &
        \begin{tabular}[c]{@{}l@{}}Take Blood\\ Sample\end{tabular} & \begin{tabular}[c]{@{}l@{}}Pre\\ Analysis\end{tabular} & \begin{tabular}[c]{@{}l@{}}Centri-\\ fugation\end{tabular} & \begin{tabular}[c]{@{}l@{}}Post\\ Analysis\end{tabular} & \begin{tabular}[c]{@{}l@{}}Inform\\ Patient\end{tabular} \\ \hline
    Duration    & 20 & 5 & 15 & 10 & 45 & 20 \\ \hline
    \begin{tabular}[c]{@{}l@{}}Role\\ Duration\end{tabular}    & 1 & 5 & 15 & 1 & 45 & 20 \\ \hline
    Cost        & 1 & 10 & 40 & 1 & 90 & 40 \\ \hline
    Location    & Waiting Room & 
            \begin{tabular}[c]{@{}l@{}}Treatment\\ Room\end{tabular} &
            Laboratory &
            Laboratory &
            Laboratory &
            \begin{tabular}[c]{@{}l@{}}Consulting\\ Room\end{tabular} \\ \hline
    \begin{tabular}[c]{@{}l@{}}IT\\ Services\end{tabular} &
            \begin{tabular}[c]{@{}l@{}}Patient\\ Database\end{tabular}  & & \begin{tabular}[c]{@{}l@{}}Diseas\\ Identification\end{tabular} & & \begin{tabular}[c]{@{}l@{}}Diseas\\ Identification\end{tabular} & \\ \hline
\end{tabular}
\caption{Data element values}
\label{tab:exampleValues}
\end{table}
% Room Names from https://www.macmillandictionary.com/thesaurus-category/british/rooms-and-departments-in-hospitals-and-clinics und https://www.nurdiono.com/name-and-function-of-wards-or-rooms-in-hospital.html

\subsection{Visual Styles}
Within the 3DViz framework three different visual styles are applied. The first is \emph{positioning} of \emph{Nodes} the second \emph{scaling} of \emph{Nodes} and the third \emph{labeling} of \emph{Nodes}. Each \emph{Node} is positioned on the X,Y and Z axis. This allows for integrating more information into the process model representation. The second style uses scaling on \emph{Nodes} and can as well be applied on the X,Y and Z axis. The third approach uses the fact that no longer a rectangle is drawn but a cube. Each side of the cube can be labeled with its own text.For example, Fig. \ref{fig:Example3DMultiView} \:\kreis{3} shows the \emph{Name} of a \emph{Node} on the front and the \emph{ID} on top of the cube.

\begin{figure}[tbh!]
	\centering
	\includegraphics[width=1\linewidth]{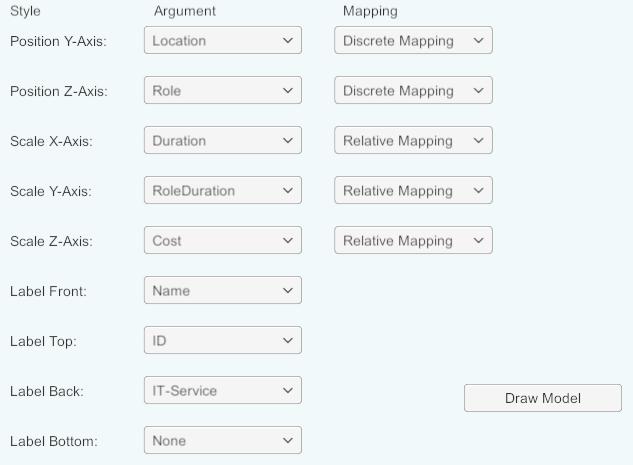}
	\caption{User interface for node representation.}
	\label{fig:UserInterface}
\end{figure}

Applying all combinations of the styles \emph{scaling} and \emph{positioning} on a process with control-flow perspective allows us to integrate five attributes within one process ($\mapsto$ RQ 1) plus additional information as a result of the cubes \emph{labels}. In the case a \emph{Node} does not contain a certain data element the \emph{Nodes} style in terms of \emph{scaling} and \emph{positioning} will be set to a default value. To deal with conflicting visual styles i.e. representing \emph{Cost} and \emph{Role} on the Z-Axis, a UI was created to resolve such conflicts manually. The UI allows for one argument per visual style. Arguments are not limited to be visualized by one style. The 3D Viz framework allows for selecting one argument for multiple styles for example, \emph{Cost} could be used on Y and Z Axis scaling. Fig. \ref{fig:UserInterface} shows the UI with many filled in values. Every row can be seen as a tuple containing a visual style, the attribute (argument) and a mapping. The mapping describes how the argument is prepared in order to be visualized. Mappings will be covered in more detail in the next section. 

In the case one tuple is selected the resulting process model shows a single attribute representation. Fig. \ref{fig:BPMNto3DModel} \:\kreis{2} depicts the example process with only using the tuple from the second row of Fig. \ref{fig:UserInterface} visual style \emph{position Z-Axis}, attribute \emph{Role} and \emph{Discrete} mapping. In the case more than one tuple is filled out we call this a multi attribute representation. Fig. \ref{fig:Example3DMultiView} depcits a 3D process model containing all the tuples from Fig. \ref{fig:UserInterface}.

\iffalse
In Fig. \ref{fig:UserInterface} more then one tuple is filled out such a case is called multi attribute representation. The resulting 3D Process model for these selected tuples is depcited in Fig. \ref{fig:Example3DMultiView}

in the case more then one tuple consisting of style, argument and method is used we call this a multiview representation. In such a case multiple views are containt within one representation.

The filled in values shown in Fig. \ref{fig:UserInterface} will be used for the multi view representation in the next sections.

filled in values for visual style \emph{position Z-Axis} for the argument \emph{Role} with the mapping \emph{Aggregation}.

\ref{fig:BPMNto3DModel} \:\kreis{2}
\fi

\subsection{Data to Visualize}
So far 3DViz consists of multiple \emph{Nodes} connected by edges visualizing the control flow view. When modeling a process more information is integrated into the model. Each \emph{Node} can have multiple arguments for example, roles, resources and data elements. These arguments need to be processed before they can be used within the representation. We differentiate 3 processing strategies, \emph{Direct Mapping}, \emph{Relative Mapping} and \emph{Discrete Mapping} ($\mapsto$ RQ 2).

\begin{figure}[tbh!]
	\centering
	\includegraphics[width=1\linewidth]{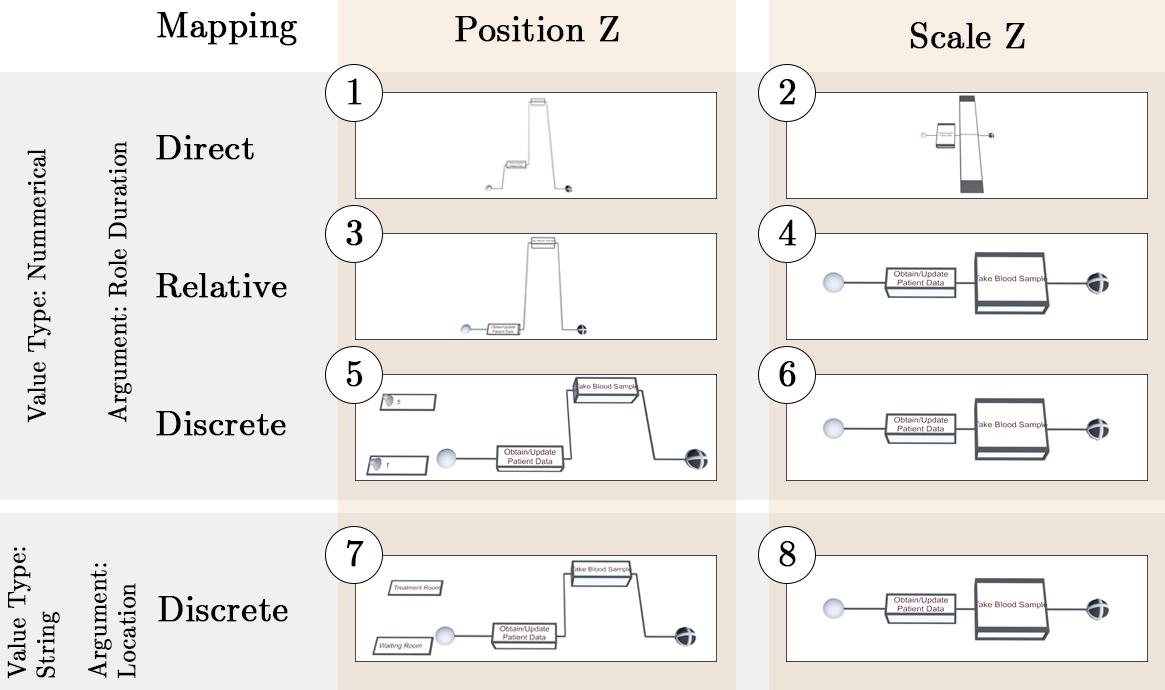}
	\caption{Mappings applied on different argument types with different styles.}
	\label{fig:StyleMappingArguments}
\end{figure}

Figure \ref{fig:StyleMappingArguments} depicts examples for all combinations of visual styles, mappings and arguments. To simplify the figure only the first two \emph{Nodes} \emph{Obtain/Update Patient Data} and \emph{Take Blood Sample} from the example process are visualized. Further
the styles \emph{position Y}, \emph{scale X} and \emph{scale Y} are not depicted as they are similar to the provided styles.

Creating a mapping for arguments imposes certain challenges such as how to handle arguments with non numerical values. Such arguments can not be used for calculating minimum and maximum values which is necessary for the \emph{Relative Mapping}. We allow the following mappings for data elements that consist of numerical values.

\begin{itemize}
\item \emph{Direct Mapping} \:\kreis{1}, \:\kreis{2} uses the values of the arguments without further modification. For example, if a value is 5 a \emph{Node} gets scaled 5 times. This mapping might not be of use in many application scenarios, but it can help to detect outliers.
\item \emph{Relative Mapping} \:\kreis{3}, \:\kreis{4} calculates a percentage value for each \emph{Node} where the minimum is 0\% and maximum is 100\%. For example, a \emph{Node} with a calculated value of 20\%  is scaled up by 20\%. Up scaling is used instead of down scaling to prevent a node getting invisible when scaled down by 100\%. 
\item \emph{Discrete Mapping} for numerical values \:\kreis{5} use multiple swim lanes and interpolate each argument to fit a specific swim lane. For scaling \:\kreis{6} each discrete group is represented by its own scaling factor ranging from 0\% to 100\%.

%For example, values 1,2,3,5,8,10 are input arguments. The users specifies a classification with 3 swim lanes. Lane 1 consists of 1, 2, 3, Lane 2 of 5 and lane 3 of 8 and 10. An other approach for aggregation is that the user defines which values shall be represented in which swim lane.
\end{itemize}

When the argument to represent is a non numerical value like resources and roles the 3DViz framework does not allow for using \emph{Direct Mapping} and \emph{Relative Mapping}. Such values can only be used with the \emph{Discrete Mapping}.

\begin{itemize}
    \item \emph{Discrete Mapping} for non numerical values \:\kreis{7} represents the arguments values in distinct swim lanes similar to numerical swim lanes. Scaling discrete non numerical values \:\kreis{8} is done by using a scaling factor ranging from 0\% to 100\% per group 
\end{itemize}

%Other arguments like resources, roles and data elements with strings do not allow for the same mappings as numerical values. These arguments are mapped onto swim lanes. For example, each role is represented by its own swim lane. If the user specifies which data elements shall be aggregated these elements can be aggregated into one swim lane.

%Aggregation of such data elements is possible if the user specifies which data elements shall be aggregated.

\iffalse
0) What do we want our views to show?

Resources (actors/roles)
    have absolute positions and are not aggregated (meaning 2 roles are 2 roles) they are not grouped together not even if they are very similar.

data elements (numbers)
    can have multiple types of positions/scalings:
        aggregation e.g. values range from zero - 10 then groups are built
            0-0.99, 1-1.99,2-2.99, ...
        absolute values zero - 10 and position on their specific values
        
        relative values zero - 10 and position percentage based

data elements (others e.g. strings, dates ...)
    have absolute positions and are not aggregated (like resources)
\fi

\subsection{Representing Single and Multiple Attributes in 3DViz}
Visual styles i.e. positioning and scaling are important when one representation contains multiple attributes. These visual styles allow to express each attribute in a unique way. When integrating multiple attributes a unique representation for each attribute is important to be able to distinguish one attribute from another.

% x = controll flow
% y = control flow - müsste auch dann gehen da sachen ja auf lanes positioniert sind

% +z
% scaling(x,y,z)

%Before these visualization possibilities can be used in a meaningfull way the data elements that shall be expressed have to be explored.

\subsubsection{Single Attribute Visualization}
Single attribute representation extends the control flow with one attribute. In case the attribute uses a data-element this element is connected to at least one task. Within the 3DViz framework the tasks are not connected to the arguments. The arguments values are used for positioning and scaling. Omitting the edges for data elements leads to a reduction of visual clutter. This will be even more important when dealing with multiple attributes.

\begin{figure}[tbh!]
	\centering
	\includegraphics[width=1\linewidth]{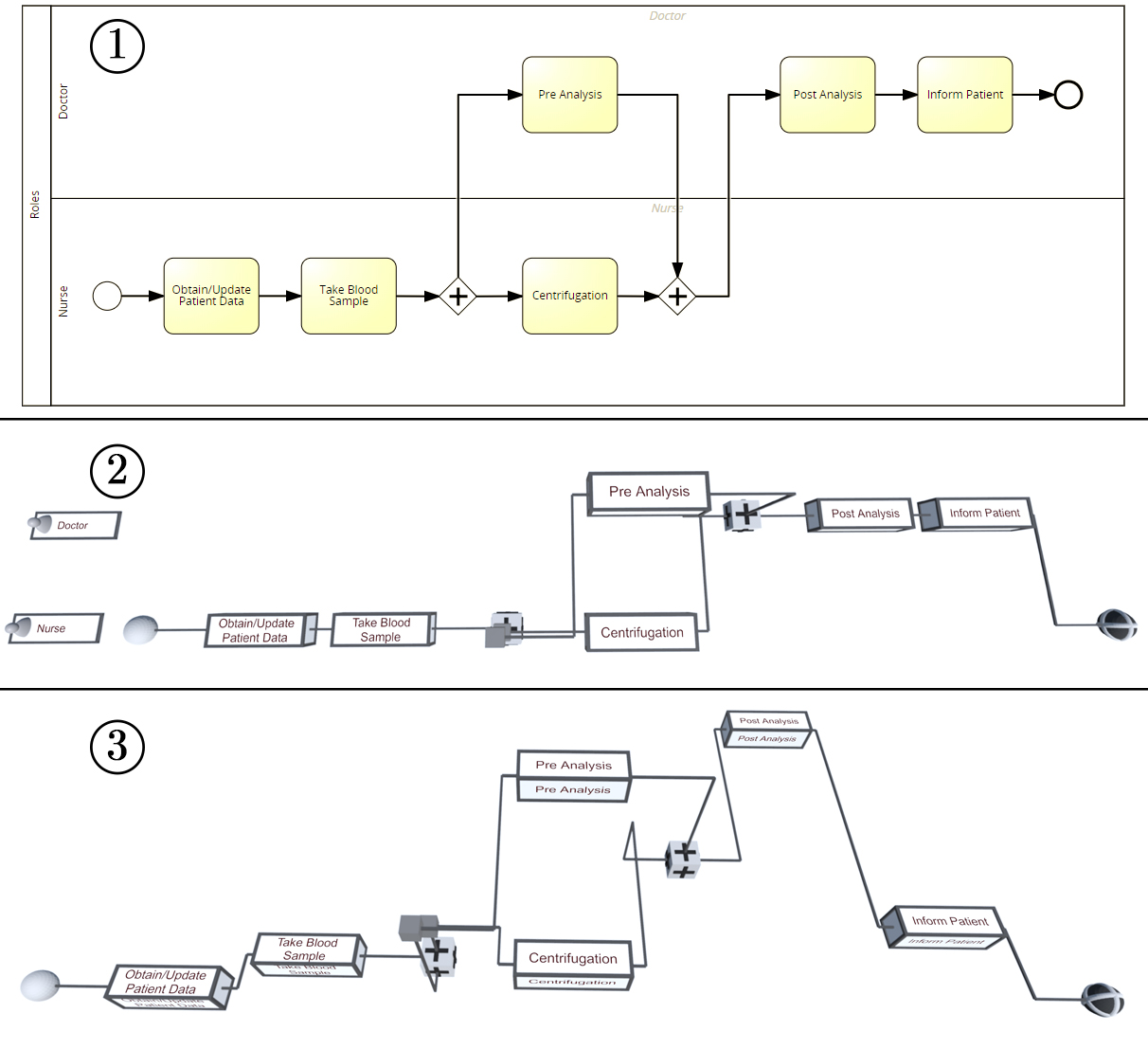}
%	\caption{1, 2 show the argument \emph{role} with \emph{Direct Mapping} on the \emph{Z-Axis}. 3 shows the \emph{Cost} with \emph{Relative Mapping} on the \emph{Z-Axis}.}
	\caption{\raisebox{.5pt}{\textcircled{\raisebox{-.9pt} {1}}}
,  \raisebox{.5pt}{\textcircled{\raisebox{-.9pt} {2}}}
 show the argument \emph{role} with \emph{Direct Mapping} on the \emph{Z-Axis}. \raisebox{.5pt}{\textcircled{\raisebox{-.9pt} {3}}}
 shows the \emph{Cost} with \emph{Relative Mapping} on the \emph{Z-Axis}.}
%	\caption{\kreis{1}, \: \kreis{2} show the argument \emph{role} with \emph{Direct Mapping} on the \emph{Z-Axis}. \:\kreis{3} shows the \emph{Cost} with \emph{Relative Mapping} on the \emph{Z-Axis}.}
	\label{fig:BPMNto3DModel}
\end{figure}

Figure \ref{fig:BPMNto3DModel} shows the same process with \emph{Signavio} \emph{Role} attribute on top \:\kreis{1}. The 3DViz visualization with style \emph{Position Z}, \emph{Discrete Mapping}, for the attribute \emph{Role} in the middle \:\kreis{2}. By showing one attribute the 3DViz Framework does not convey its full potential. In example \:\kreis{2} the angle from top was chosen to show how similar those two representations are.

Figure \ref{fig:BPMNto3DModel} \:\kreis{3} depicts another single attribute. Representing the style \emph{Position Z}, \emph{Relative Mapping} for the attribute \emph{Cost}. This approach differentiates from other approaches suggested by literature \cite{finkelstein1992viewpoints,BeckerGuid} as the arguments values changes the \emph{Nodes} Z-Axis position. One can spot that the \emph{Nodes} \emph{Pre Analysis} and \emph{Post Analysis} are positioned in the back as those are the \emph{Nodes} with highest cost.

%\begin{figure}[tbh!]
%	\centering
%	\includegraphics[width=1\linewidth]{images/BPMNto3D.jpg}
%	\caption{Both models show the same process, on the left a BPMN model and on the right our 3D model}
%	\label{fig:BPMNto3DModel}
%	\vspace{-0.2cm}
%\end{figure}

\subsubsection{Multiple Attribute Visualization}
Limiting one representation to a single attribute aggravates the ability to explore connections between attributes. Shifting from one attribute to another will not allow to recognize these connections either. With 3DViz we allow for combining multiple attributes into one representation ($\mapsto$ RQ 1). Within a 2D setting comparing arguments can be accomplished by switching between representations. In a 3D representation the arguments can be included within one representation and compared i.e. by the size of a cube. Further differences when comparing within a 3D representation compared to a 2D representation is that one does not lose focus on the whole picture of the process.

When using multiple attributes we suggest to use non numerical arguments e.g., resources and roles with a \emph{Discrete Mapping} for the positioning on the Y or Z-Axis. While numerical values should be used for scaling of nodes.

\begin{figure}[tbh!]
	\centering
	\includegraphics[width=1\linewidth]{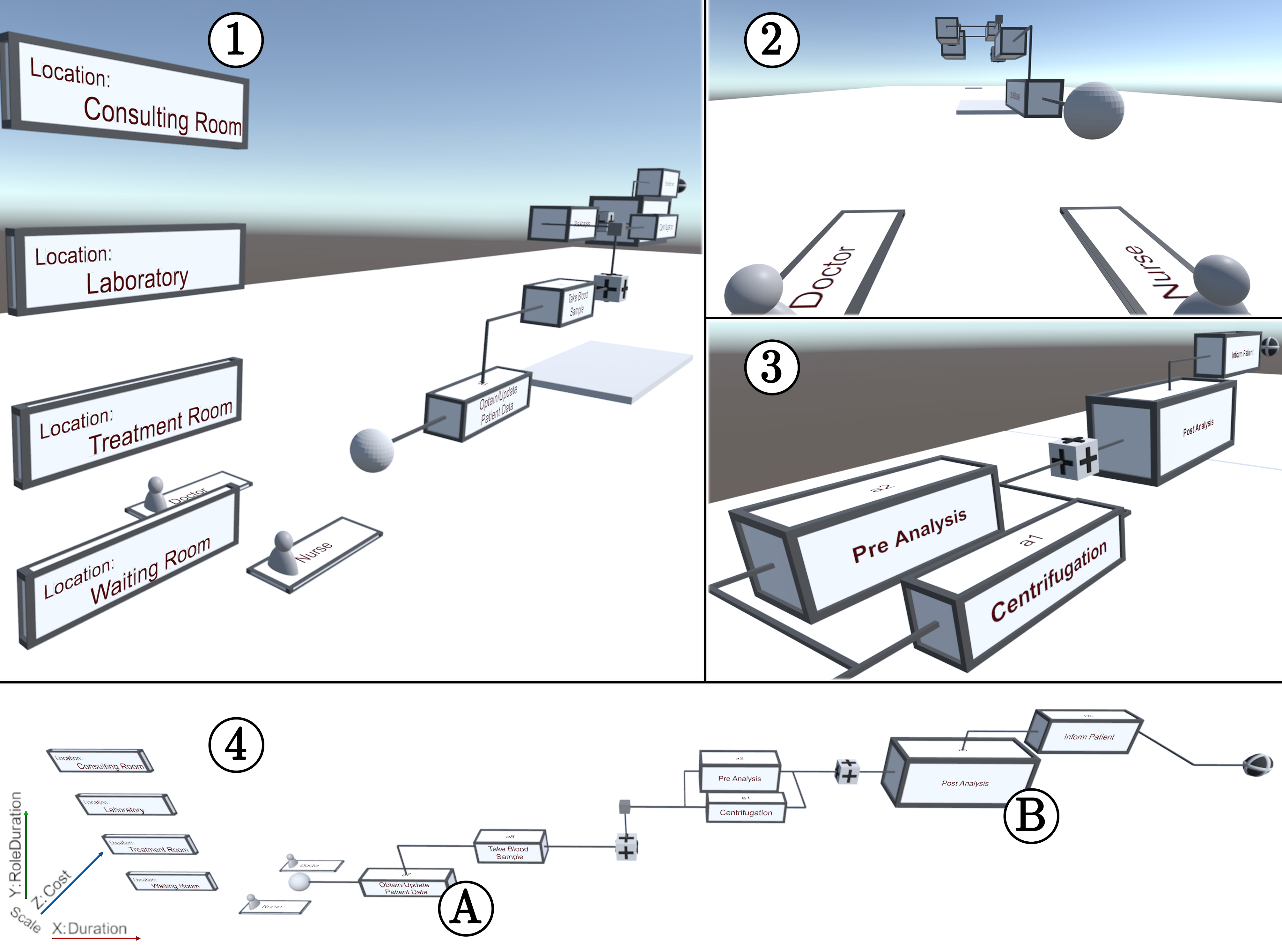}
	\caption{Different angles on the 3D representation of the centrifuge process.}
	\label{fig:Example3DMultiView}
\end{figure}

Figure \ref{fig:Example3DMultiView} shows the example process with the mappings from Fig. \ref{fig:UserInterface}, \emph{positioning} for \emph{location} \:\kreis{1}, and \emph{roles} \:\kreis{2}, \emph{scaling} \:\kreis{3} for \emph{duration}, \emph{role duration} and \emph{cost} and \emph{labels} for \emph{id}, \emph{name} and \emph{IT-Service}. On the bottom of Fig. \ref{fig:Example3DMultiView} \:\kreis{4} gives an overview of the whole process with the scaling axis depicted on the bottom left. Within the framework the scaling axis is depicted every time a scaling is active.

Based on the 3D representation depicted in Fig. \ref{fig:Example3DMultiView} the following conclusions can be made and can be part of an in depth analysis for potential process improvements.  Two \emph{Nodes} \emph{Pre Analysis} and \emph{Centrifuge} are performed in parallel at the same location by two different \emph{Roles} \:\kreis{3}. The duration of a \emph{node} is not responsible for a \emph{Nodes} \emph{Cost} \:\kreis{4}\:\kreis{A},\:\kreis{B} whereas, the \emph{Role duration} and \emph{Cost} seem to be related \:\kreis{3} e.g. \emph{Nodes} \emph{Pre Analysis} and \emph{Post Analysis}. Under consideration of the arguments \emph{Cost} and \emph{Role duration} \emph{Doctors} are more expensive than \emph{Nurses}.

%\todo{Neue implementierungsmöglichkeit: Use original borders to show how much something increased, to give a better indicator for comparison, von der farbe her ein schwächeres grau verwenden. Aufpassen das parent object muss der cube sein.}

The example combines five attributes in one representation. For inexperienced users this can be overwhelming. However, the example demonstrates the full potential of 3DViz. The comprehension one can obtain from a 3DViz representation depends on the complexity of the process model that is visualized and the level of experience a user has with 3D representations.

In conclusion we want to emphasize that representing multiple attributes within one model is a suitable approach to detect relationships between attributes. 
%However these models can get overwhelming and therefore using a 3D scene can ease orientation and navigation within the representation specially when wearing a HMD. 

\section{Evaluation}
\label{sec:Evaluation}
The evaluation consists of two parts: a user survey and an evaluation on the applicability of the 3DViz framework. The user survey outlines how a 3D process representation is perceived and where the strengths of such a representation can be found. To show the applicability of the 3DViz framework the framework was implemented and a performance analysis carried out.

\subsection{User Survey}
The first goal of the survey was to evaluate if 3D representations are applicable for business process representation. The second goal was to identify the strengths of a 3D process visualization.

\subsubsection{Survey Design}

%Target Audience
The survey was designed to be conducted during a public event called \emph{Long Night of Science}. During the event the attendees participated in a demonstration of a physical process were goods had to be moved from the storage unit to a consumer. During the execution of the process the attendees were introduced to 3DViz. While they executed the physical process they were able to see and explore the 3D process model.

Based on this experience the participants answered a short survey containing two demographic questions and six 3DViz related questions. The survey was designed in accordance with the \emph{Ten Commandments} \cite{Porst}. A pretest with four participants, two familiar with the topic and two not familiar with the topic was used to evaluate and improve the question design.

\subsubsection{Sample}
\begin{figure}[tbh!]
	\centering
	\includegraphics[width=1\linewidth]{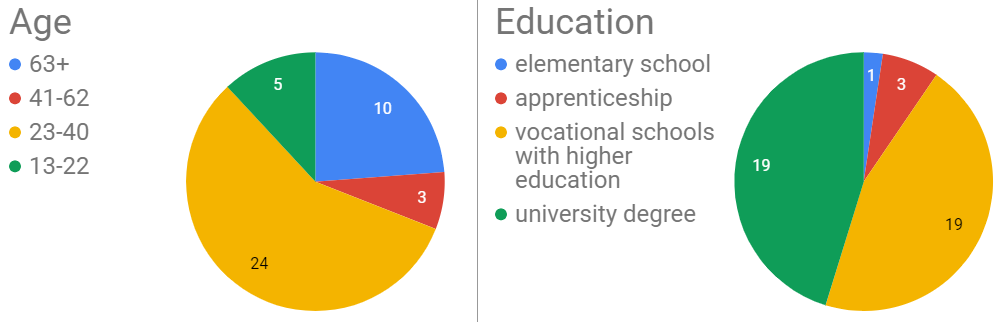}
	\caption{Age and education of the survey participants.}
	\label{fig:EducationEval}
\end{figure}

By reaching out during a public event we wanted to reach a diverse set of people. In total 42 participants answered the survey. Figure \ref{fig:EducationEval} depicts the distribution for age and education among the participants. The age distributes as followed, 10 (23.8\%) participants are older than 63 years, 3 (7.1\%) participants between 41-62 years, 24 (57.1\%) participants between 23-40 years, and 5 (11.9\%) participants between 13-22 years.
 None of the participants was younger than 13 years. The education is biased towards higher educations such as vocational schools with higher education and university degree each selected by 19 (45.2\%) participants. The remaining 9.6\% distribute to apprenticeship 3 (7.2\%) and elementary school 1 (2.4\%).

\subsubsection{Results and Discussion}
The first question the participants had to answer was if they think 3D representations can be useful when representing a process model. In total 38 (92.8\%) of the participants answered with yes while 3 (7.2\%) answered with no. This indicates a strong sign that 3D representations can be a benefit for process model visualization.

\begin{figure}[tbh!]
	\centering
	\includegraphics[width=0.7\linewidth]{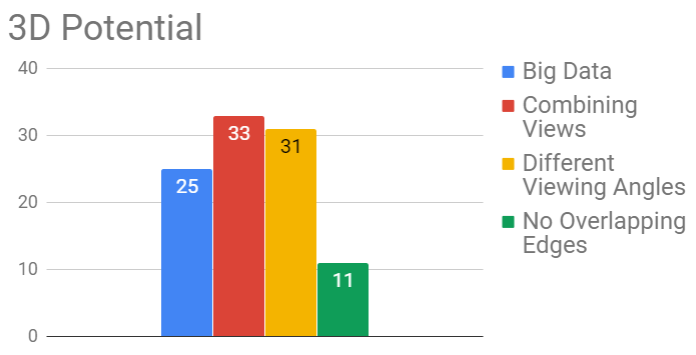}
	\caption{Potential areas where 3D process visualization can thrive.}
	\label{fig:ArgumentData}
\end{figure}

The next question should reveal the areas a participant identifies potential for 3D representations. The survey provided four check boxes and one open form to fill out. Figure \ref{fig:ArgumentData} summarizes the results. From the 42 participants 33 (80.5\%) marked that combining multiple attributes into one representation is a good application for 3D process models. 31 (75.6\%) participants found that changing the viewing angle and rotating around the process model is suitable for 3D applications. 25 (61\%) of the participants think that 3D representations can be used to express big data. Surprisingly only 11 (26.8\%) think that 3D representations are a good fit to represent less overlapping edges. Three Participants contributed to the open question. For general readability the statements are translated from German to English. The first two statements "Use VR for exploration" and "Use AR and VR technology" are covered within this paper. In our demo show case we did not have the possibility to demonstrate 3DViz with such devices. The third statement "Use visually different paths through the process." was so far not under consideration for the 3DViz framework. However, we argue that for the current implementation differentiation between paths is not necessary. Such visually different paths could be interesting when considering process mining or run-time analysis.

The last question of the survey was if using a 3D scene is a good solution to increase orientation within a 3D representation. Answered by 40 participants (97.6\%) with yes it is clear that using a 3D scene is a very popular choice.

In summary we identified that 3D representations are suitable for process representations. Within a 3D representation there are areas such as attribute combination that can benefit from utilizing the third dimension. When representing a 3D process model a scene around the process was very well received by the participants. Such a scene aids navigation and orientation within a 3D representation.

%The understanding of a 3D process is improved by allowing the user to interact with the model by rotating the point of view. 

\subsection{Applicability}
To demonstrate the applicability of 3D process attribute representation the approach has been prototypically implemented with the Unity framework. The prototype is called 3DViz and available here\footnote{\label{foot:url}\url{http://gruppe.wst.univie.ac.at/projects/crisp/index.php?t=visualization}}.

\begin{figure}[tbh!]
	\centering
	\includegraphics[width=0.7\linewidth]{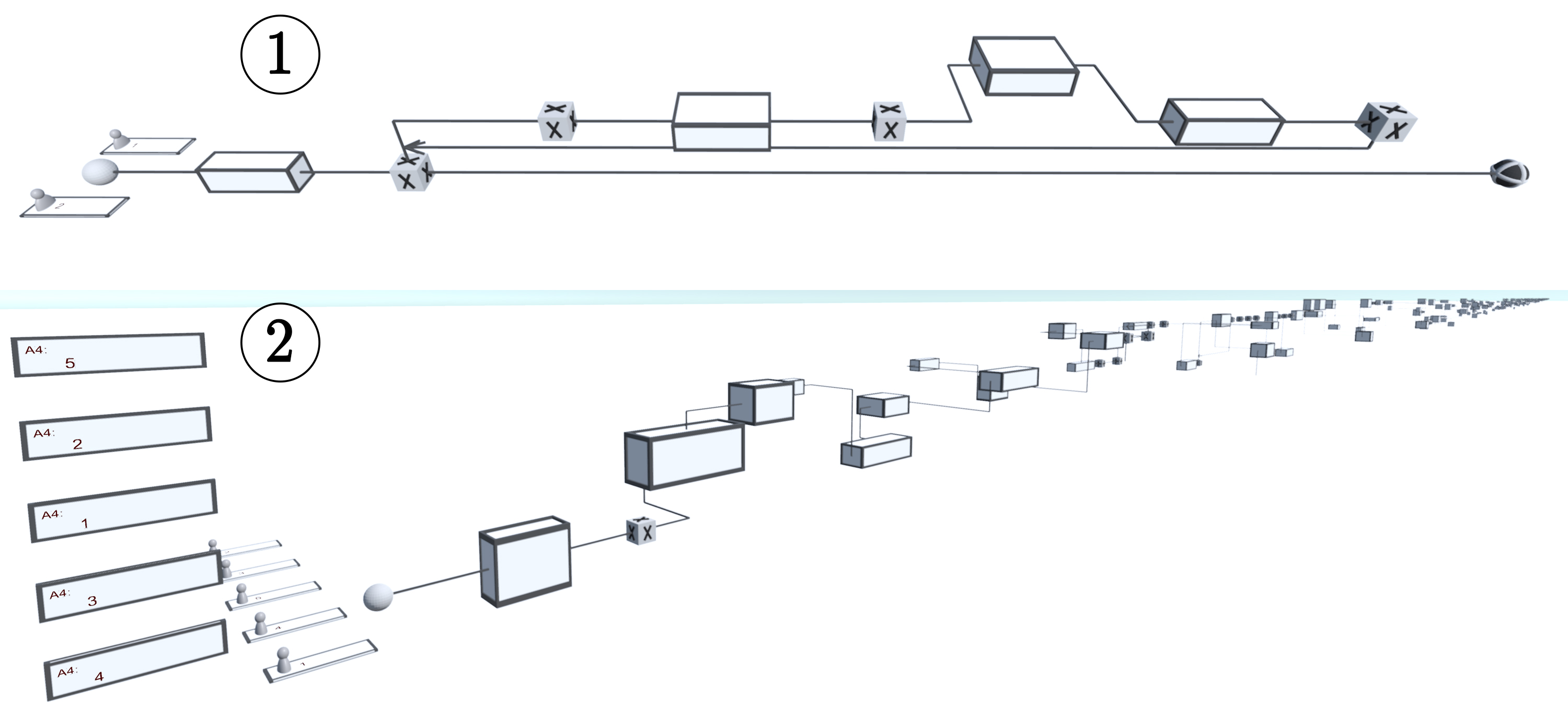}
	\caption{Two randomly created process models the first with 4 \emph{Nodes}, 2 control-flow elements, and 2 arguments, the second with 1024 \emph{Nodes}, 512 control-flow elements and 5 arguments. (label unreadability intended)}
	\label{fig:RandomExample}
	\vspace{-0.2cm}
\end{figure}

We randomly created process models with a tool available here$^1$. The tool allows to specify the amount of \emph{Nodes} that shall be created, the amount of arguments and the amount of control-flow related elements such as \emph{Parallel}, \emph{Xor}, and \emph{Loops}. The arguments and control-flow related elements are randomly assigned. Figure \ref{fig:RandomExample} \:\kreis{1} depicts an example with 4 \emph{Nodes}, 2 arguments, and 2 control-flow elements as input parameters. One of the arguments is applied with discrete mapping on the Z-Axis position the other one with relative mapping on the Z-Axis scaling. Resulting in a process model starting with a \emph{Node} followed by a \emph{Loop} that contains 3 \emph{Nodes} and an \emph{Xor} one of the \emph{Nodes} consists of a different argument and is represented on an other swim lane. For demonstration of a large process Fig. \ref{fig:RandomExample} \:\kreis{2} depicts an example with 1024 \emph{Nodes}, 512 control-flow elements and 5 arguments one applied per visual style.

\begin{figure}[tbh!]
	\centering
	\includegraphics[width=0.7\linewidth]{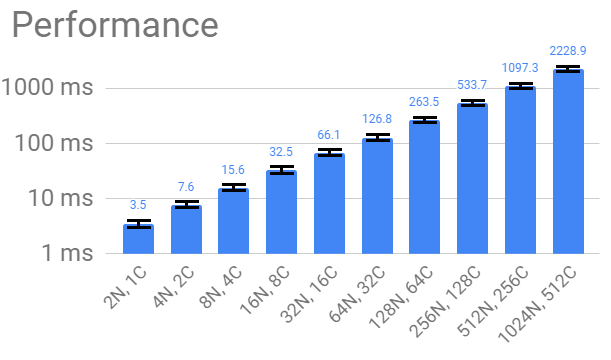}
	\caption{Perfromance analysis of 3DViz. X-Axis depicts \emph{Nodes} and control-flow elements. Y-Axis depicts the time in miliseconds needed to visualize the process model.}
	\label{fig:Performance}
\end{figure}

For the performance analysis the time between receiving the file from the random process creation tool and drawing of the process is finished is measured. The analysis is carried out on a mid to high end PC with quad-core processor with 3.60 GHz, 16 GB RAM, and a GTX 1060 graphics card. Figure \ref{fig:Performance} depicts the measurements ranging from 2 \emph{Nodes} and 1 control-flow element depicted as \emph{2N, 1C} up to 1024 \emph{Nodes} and 512 control flow elements \emph{1024N, 512C}. All of the measurements are performed with five arguments. Per group 10 measurements were taken. The average time needed is displayed on the Y-Axis of Figure \ref{fig:Performance} with the co-efficient of variation displayed on top of each bar. The co-efficient of variation is decreasing from 15\% \emph{2N,1C} towards 3.5\% for \emph{1024N, 512C}. Figure \ref{fig:Performance} shows a linear growth demonstrating that the framework is applicable for small and huge process models.

Based on the evaluation results we argue that the applicability is given. 3DViz is capable of displaying models containing 1024 \emph{Nodes} and 512 control-flow elements within 2 seconds.

\section{Conclusion and Outlook}
\label{sec:con}

The proposed 3DViz framework allows to represent multiple attributes such as resources, data, and roles within one 3D representation. Representing multiple attributes at once allows for finding relations between attributes. Such a representation shows a lot of information therefore we propose to use modern technology like head mounted displays to explore 3D representations. When wearing such a device orientation can be difficult. Therefore we propose to use a 3D scene to aid orientation when exploring a 3D process model.

3DViz was evaluated with a user study and applicability evaluation. The studies show that 3D process attribute representation is a good solution for representing multiple attributes. Performance wise the evaluation showed a linear growth and that the framework is capable of representing huge process models within seconds. 

For future work we want to expand towards representing runtime-execution and compliance violations within 3DViz.

%\subsubsection{Acknowledgment} {This work has been funded by the Vienna Science and Technology Fund (WWTF) through project ICT15-072.}

\bibliographystyle{splncs03}
\bibliography{references}

\end{document}